# Topological transition in spiral elastic valley metamaterials

Shuaifeng Li, Jinkyu Yang

Department of Aeronautics and Astronautics, University of Washington, Seattle, Washington, 98195-2400, USA

Elastic valley metamaterials offer an excellent platform to manipulate elastic waves and have potential applications in energy harvesting and elastography. Here we introduce a series of strategies to realize topological transition in spiral elastic valley metamaterials by parameter modulations. We show the evolution of Berry curvature and valley Chern number as a function of inherent parameters of spiral, which further results in a general scheme to achieve topological valley edge states. Our strategy leverages multiple degrees of freedom in spiral elastic valley metamaterials to provide enhanced opportunities for desired topological states.

## I. INTRODUCTION

Topological valley metamaterials have emerged as a remarkable impact not only on condensed matter physics but also on manipulation of waves in electronics, photonics and phononics [1–9]. Wave propagation in topological valley metamaterials possesses prominent applications in information carrier and light modulation with its remarkable feature: the robust valley-polarized transport [1–4,6–8,10,11]. Therein, topological transition is a necessary process in metamaterials to realize topological states. Several strategies have been proposed to achieve the topological transition. For example, the inherent degree of freedom, such as adjusting the geometric parameters, can be exploited to invert the topological phase of metamaterials [4,6,9,12]. Besides, the external degree of freedom, such as deformation, can also affect the band inversion dynamically [13,14]. However, existing metamaterials are still in lack of the control knob to generate the desired topological phase at will. One of the reasons is that the inherent simple structures are insufficient to manipulate the geometric parameters.

Recently, an elastic valley metamaterial that consists of a hard spiral in a soft hexagonal matrix has been proposed as a new design to achieve the valley topological insulators [9]. On the contrary to conventional valley systems, the spiral elastic valley metamaterials have the high complexity of internal structure, which has a series of design parameters to manipulate. For example, this spiral system possesses an inherent chiral anisotropy, which results in valley anisotropy and the exceptional Berry curvature distribution. Based on this, the elastic topological valley edge state relying on frequency has been demonstrated to have potential applications in signal processing and frequency selector.

Though this system has shown promising properties and improved controllability in comparison to the symmetric valley system, there remain several natural questions about this spiral valley system. For example, since this is a chiral anisotropic system, how does the chirality affect the topological properties? How does the topological transition happen in an asymmetric system? Addressing these challenging questions is meaningful to understand and enrich the intrinsic physics of topological valley states.

In this manuscript, we report topological transition via parameter modulations in the spiral elastic metamaterials. We discuss the effect of chirality, rotation angle, number of turns and thickness of the spiral to the topological properties in terms of Berry curvature and valley Chern number. Topological transition process of spiral elastic metamaterials is clearly shown by the inversion of the Berry curvature and valley Chern number. After the key factors contributing to the topological transition are determined, we propose a general scheme to generate the desired topological phase of these elastic metamaterials. Furthermore, we demonstrate the topological valley edge states by exploiting this general scheme. Thereby, our strategy provides enhanced degrees of freedom and opportunities to achieve topological transition by leveraging the spiral architecture's parameter modulations.

## II. TOPOLOGICAL TRANSITION VIA PARAMETER MODULATIONS

FIG. 1 shows the schematic of the spiral elastic valley metamaterial considered in this study. The hard spiral made of polylactic acid (PLA) is embedded as a scatterer in the triangular unit cell matrix made of soft hydrogel. The mechanical properties of the spiral PLA are: mass density 1250 kg/m$^3$, Young's modulus 2.1 GPa, and Poisson's ratio 0.36. The mechanical properties of soft hydrogels are: mass density 1000 kg/m$^3$, Young's modulus 18 kPa, and Poisson's ratio 0.5. Four parameter modulations are shown around the original unit cell of the spiral elastic metamaterial, where the side of hexagon $c$ = 14 mm and initial spiral radius $a_i$ = 1.5 mm. Modulations of chirality $C$, rotation angle $\theta$, number of turn $n$, and thickness of spiral $d$ are illustrated counterclockwise. For simplicity, we denote our spiral elastic metamaterial by ($C$, $\theta$, $n$, $d$ [mm]) to discuss the topological transition. All simulations are done under the plane strain condition using Comsol Multiphysics.

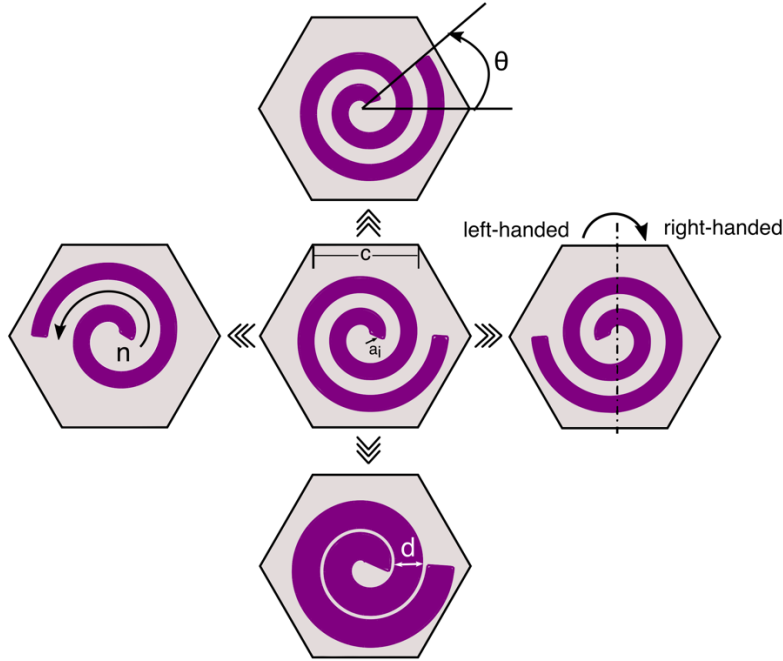

FIG. 1. Schematics of the spiral elastic valley metamaterials. The scatterer (Archimedean spiral structure) is shown in purple and the soft matrix is shown in beige. Modulations of chirality, rotation angle, number of turn, and thickness of scatterer are illustrated on the right, top, left, and bottom, respectively.

In the first place, we investigate the effect of rotation angle of spiral around the center of the hexagon ($\theta$) to the topological properties. According to our previous work, there exists a band inversion under certain circumstance between the second and the third band [9]. Therefore, we focus on these two bands of (right-handed, $\theta$, 2, 2). We calculate two bands around the K point when the spiral is rotated around the center from 20° to 40° with the increment 1°. The blue line in FIG. 2(a) represents the maximum frequency of the second band and the red line represents the minimum frequency of the third band. The enclosed area, which is the band gap range, experiences an open-closed-reopen process (see the narrowing bandgap around 30° region). Because of the low-order symmetry of the spiral, the band gap cannot completely close.

We calculate the valley Chern number $C_v = \frac{1}{2\pi}\int i\nabla_k \times \langle u(\boldsymbol{k})|\nabla_k|u(\boldsymbol{k})\rangle d^2\boldsymbol{k}$ based on the third band using the numerical method [15]. As shown in FIG. 2(a), the green line indicates that the valley Chern number has a sudden increase at around 29°, suggesting that our spiral elastic metamaterial has a topological transition. Within one topological phase, the valley Chern number has little fluctuation, which suggests that the rotation has little effect on the valley Chern number until it crosses the critical angle around 29°. Besides, the valley Chern number is always within (-0.5, 0.5) due to the strong spatial inversion symmetry breaking [9,16–18].

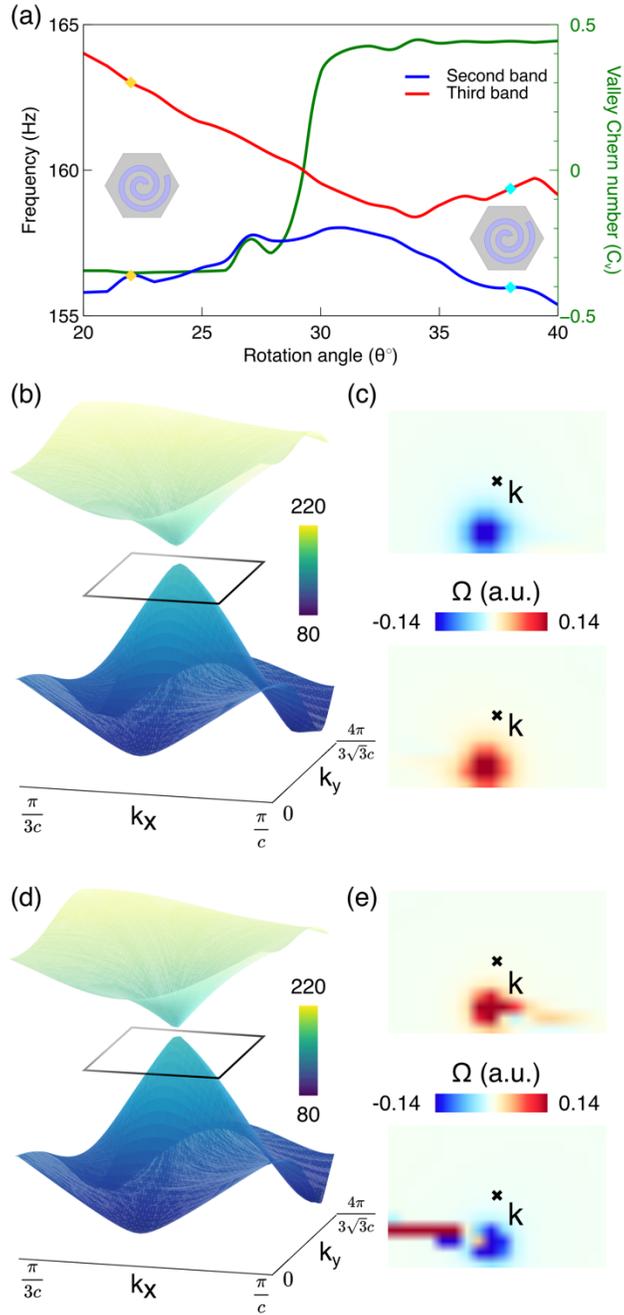

FIG. 2. (a) The band evolution as a function of rotation angle. The blue and red line represent the maximum frequency of second band and the minimum frequency of the third band, respectively, as the spiral rotates. The green line indicates the valley Chern number as a function of rotation angle. (b) Band structure of (right-handed, 22, 2, 2) elastic metamaterial, which is marked by yellow diamonds in (a). (c) The corresponding Berry curvature. Black crosses indicate the corners

of Brillouin zone (K point). (d) Band structure of (right-handed, 38, 2, 2) elastic metamaterial, which is marked by cyan diamonds in (a). (e) The corresponding Berry curvature. Black crosses indicate the corners of Brillouin zone (K point). The area for the calculation of Berry curvature is shown in (b) and (d) in black lines, which spans $\mathbf{k_x}$ from $\frac{8\pi}{15c}$ to $\frac{4\pi}{5c}$, $\mathbf{k_y}$ from $\frac{\pi}{2\sqrt{3}c}$ to $\frac{5\pi}{6\sqrt{3}c}$.

FIG. 2(b) shows the band structure around the K point of (right-handed, 22°, 2, 2) as indicated by the yellow diamonds in FIG. 2(a). The second band and the third band behave like the typical valley structure with a small band gap. The Berry curvature distributions of the corresponding bands calculated within the black lines (FIG. 2(b)) are shown in FIG. 2(c), where they have the opposite sign and the extrema of Berry curvature are deviated from the K point. This discrepancy is caused by the mismatch between the asymmetrical spiral and the triangular lattice [9]. When the spiral continues rotating to 38° until it has the configuration of (right-handed, 38°, 2, 2) [cyan diamonds in FIG. 2(c)], we notice that the band structure illustrated in FIG. 2(d) is similar to that of (right-handed, 22°, 2, 2). However, the Berry curvature distributions are completely distinct. The Berry curvature of the second band has negative values and that of the third band has positive values. This inversion of the Berry curvature indicates the band inversion and the topological transition. The band inversion often reflects on the eigen modes of metamaterials, but interestingly, the eigen modes of these two configurations at K point do not show the swap of the two bands clearly (see the APPENDIX A). This is one of unique aspects of our spiral valley system in contrast to conventional symmetric valley system.

Spiral is the natural chiral element, where the right-handed and left-handed spirals can be transformed by parity inversion [19,20]. In order to explore the effect of chirality on the topology of the bands, we fix $d = 2$ mm and choose several configurations made from combinations among chirality, rotation angle, and number of turns. Likewise, we calculate Berry curvature around K point and integrate it for the valley Chern number for all configurations. Note that the valley Chern number calculated here is based on the third band because the second band is largely coupled with the first band which will cause the inaccuracy of the valley Chern number. In FIG. 3(a), the green filled circles show the variation of valley Chern number for (right-handed, $\theta$, 1.5, 2) as a function of rotation angle. Meanwhile, green unfilled circles represent the variation of valley Chern number for (left-handed, $\theta$, 1.5, 2). Likewise, in FIG. 3(b), magenta filled triangles and unfilled triangles show the variation of valley Chern number for (right-handed, $\theta$, 2, 2) and (left-handed, $\theta$, 2, 2), respectively. From the two graphs, the rotation is able to cause the topological transition, which confirms the conclusion in FIG. 2. More notably, massive calculations of valley Chern numbers of different configurations shows that when the chirality of spiral changes, e.g., from right-handed to left-handed spiral, the sign of valley Chern number always changes, suggesting that topological transition should happen when the chirality changes. That is, we find that the Chern number data points of the left-handed counterparts (unfilled circles/triangles) can always be found on the other side of the right-handed geometries (filled circles/triangles). Note that we calculate the valley Chern number from 0° to 360° with the increment 20° but several missing data points are found because Berry curvatures of some configurations are distributed with both positive and negative values, which poses a challenge to the integral. We also observe in passing that when the number of turns $n$ changes from 1.5 to 2, the sign of valley Chern number is flipped, which has been shown in previous work [9].

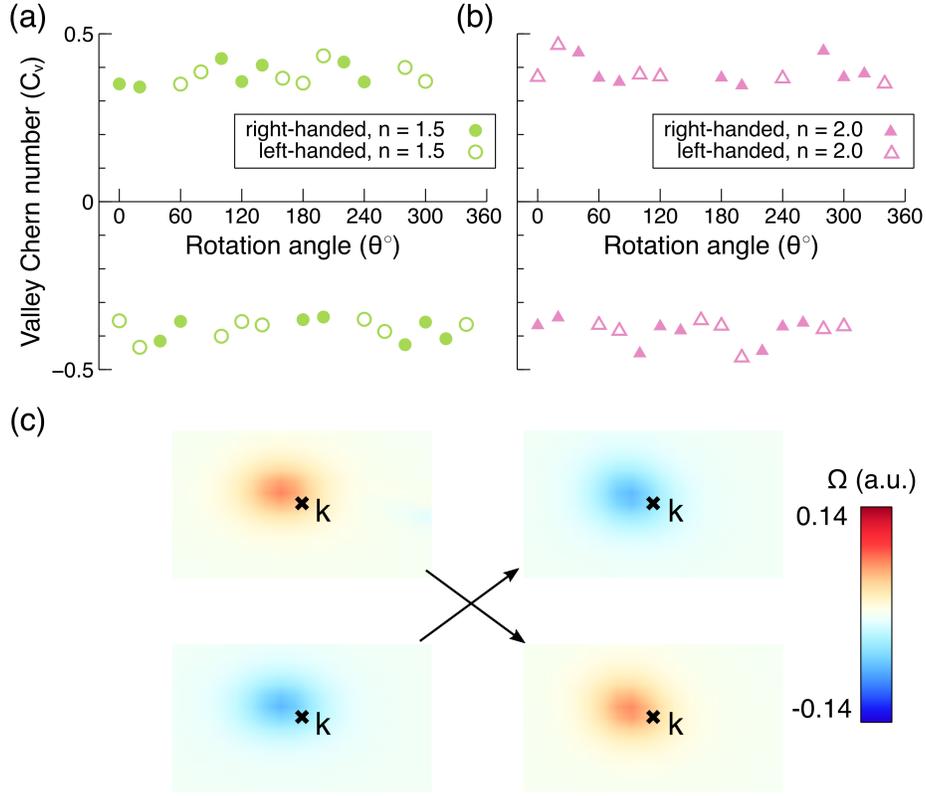

FIG. 3. (a-b) Evolution of valley Chern number for elastic metamaterials under different chirality. (a) When the rotation angle $\theta$ is varying from 0° to 360°, valley Chern numbers for (right-handed, $\theta$, 1.5, 2) and (left-handed, $\theta$, 1.5, 2) are shown in green filled circles and green unfilled circles, respectively. (b) Valley Chern numbers for (right-handed, $\theta$, 2, 2) and (left-handed, $\theta$, 2, 2) are shown in magenta filled triangles and magenta unfilled triangles, respectively. (c) The left panel shows the Berry curvature of the third and the second band of (left-handed, 60°, 1.5, 2) metamaterial, while the right panel shows that of (right-handed, 60°, 1.5, 2) metamaterial.

We then take (left-handed, 60, 1.5, 2) and (right-handed, 60, 1.5, 2) as examples to demonstrate the topological transition caused by the chirality modulation. Top row in FIG. 3(b) shows Berry curvature distributions for both cases of the third bands, while the bottom row shows that of the second bands. The Berry curvature distribution, likewise, is deviated from the K point because of the mismatch between the asymmetrical spiral and the triangular lattice. When the left-handed spiral is transformed to the right-handed spiral, there is a clear sign flip of the Berry curvature, suggesting that the topological transition happens in this process.

To continue discussing the effect of thickness of spiral to the topological properties, we calculate the valley Chern number for several elastic metamaterials with different configurations ($C$, 60°, $n$, $d$). In FIG. 4, while $d$ varies from 1 mm to 4 mm, the evolution of valley Chern number of (right-handed, 60°, 1.5, $d$) is shown in red solid line. The corresponding Berry curvature distribution is shown near the red solid line. We notice that apart from the deviation of Berry curvature, the magnitude of Berry curvature becomes smaller and the distribution becomes dispersed when $d$ increases to 4 mm, thus resulting in the decreasing valley Chern number. Since the spiral approaches a solid circle with the increase of $d$, this will reduce the broken spatial inversion, which will result in the decrease of Berry curvature.

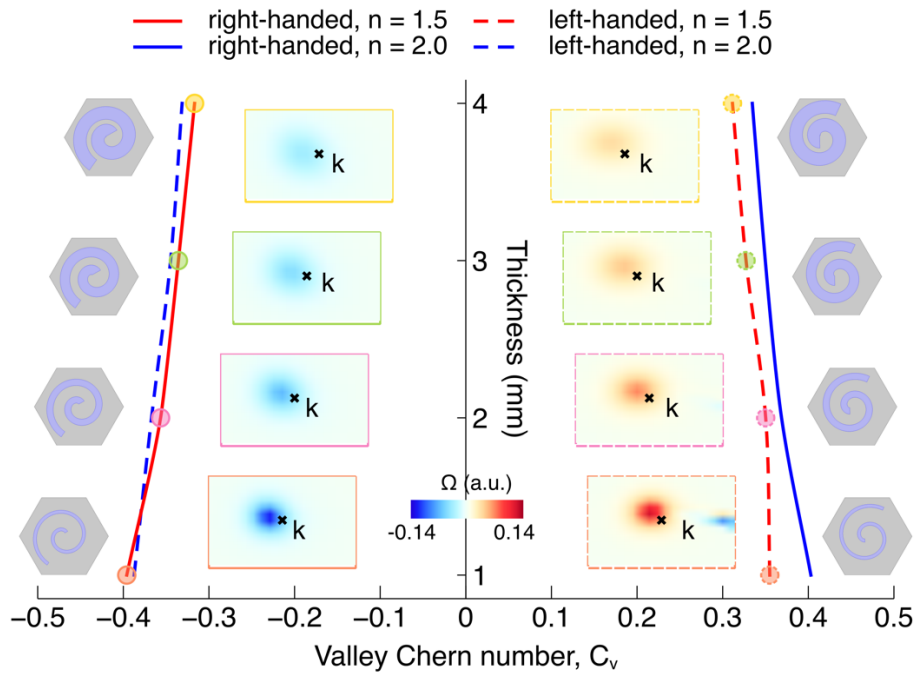

FIG. 4. Evolution of valley Chern number for elastic metamaterials under different thicknesses. When the thickness $d$ is varying from 1 mm to 4 mm, valley Chern numbers for (right-handed, 60°, 1.5, $d$), (left-handed, 60°, 1.5, $d$), (right-handed, 60°, 2, $d$), and (left-handed, 60°, 2, $d$) metamaterials are shown in red solid line, red dashed line, blue solid line and blue dashed line, respectively. The corresponding Berry curvatures of the third band are shown beside lines. The corresponding configurations are also shown beside lines.

As discussed in the last section, the change of chirality can induce the topological transition. The red dashed line represents the evolution of valley Chern number of (left-handed, 60°, 1.5, $d$), which displays the positive values. The corresponding Berry curvatures are shown beside. They have the same trend as the previous one. Similarly, we can deduce that the valley Chern number of (right-handed, 60°, 2, $d$) will be the opposite compared with that of (right-handed, 60°, 1.5, $d$), as shown in blue solid line, which has negative values. The left-handed one, on the contrary, has valley Chern numbers with positive values. We also notice that increasing $n$ from 1.5 to 2 will enhance the absolute value of valley Chern number.

According to the above analysis, we can conclude that chirality $C$, rotation angle $\theta$, and the number of turns $n$ of the spiral play an important role in topological transition. Chirality transformation causes topological transition when the parity changes. Rotation will also change the topology of bands and it is related to the symmetry of the lattice. When the rotation angle varies from 0° to 360°, there is a periodic change of topological properties, where six topological transition points exist in this process determined by the triangular lattice. We also confirm that changing number of turns results in topological transition. However, changing the thickness of spiral only adjusts the valley Chern number instead of flipping the sign. In possession of the information of topological transition via parameters modulation, we can design the desired topological phase of spiral elastic valley metamaterials.

### III. ELASTIC TOPOLOGICAL VALLEY EDGE STATES

After obtaining the general scheme of generating different topological phase, we create an elastic topological insulator based on our conclusion we draw in the last section. We can confirm that (right-handed, 60°, 1.5, 2) has a positive Berry curvature and a negative Berry curvature in its second and third bands, respectively, as observed from FIG. 3(b). To find an elastic metamaterial with the opposite topological properties and fully reflect our conclusion shown before, we choose left-handed spiral first, which will invert the topological phase once [i.e., (***right***-handed, 60°, 1.5, 2) → (***left***-handed, 60°, 1.5, 2)]. Then we are aware that rotation also induces topological transition, so $\theta = 0°$ is chosen to invert the topological phase twice [i.e., (left-handed, **60°**, 1.5, 2) → (left-handed, **0°**, 1.5, 2)]. After we select $n = 2$ as the number of turns of the spiral, we invert the topological phase three times in total [i.e., (left-handed, 0°, **1.5**, 2) → (left-handed, 0°, **2**, 2)]. Thus, we have the opposite topological phase if we choose (left-handed, 0°, 2, 2) compared to the original (right-handed, 60°, 1.5, 2). Lastly, to make sure that we have an overlapping band gap between two components, we change the thickness $d$ (see APPENDIX B that shows the frequency range of band gap as a function of $d$). Conclusively, (left-handed, 0°, 2, 1) is chosen to have the opposite topological phase and the overlapping band gap with (right-handed, 60°, 1.5, 2). Therefore, we denote generated elastic valley metamaterial as (right-handed, 60°, 1.5, 2 | left-handed, 0°, 2, 1).

FIG. 5(a) illustrates the projected band structure calculated by a sandwiched supercell (right-handed, 60°, 1.5, 2 | left-handed, 0°, 2, 1 | right-handed, 60°, 1.5, 2). The geometry of the supercell is shown in FIG. 5(b), where the interface is zigzag. In the band structure, the bulk states are shown in gray. Because two elastic metamaterials are topologically different, two topological interface states appear within the band gap as a result of the bulk-edge correspondence, as displayed in red and blue lines. We choose four points on the topological interface states located at **k** = 0.6 and **k** = 0.7. By checking the eigen displacement field at these points, FIG. 5(b) indicates the vibrations are concentrated around the interface marked by the arrows (magnified views of the interfaces are shown in the insets). As shown by the leftmost (orange) and rightmost (pink) cases in FIG. 5(b), the eigen displacement fields corresponding to the blue line in FIG. 5(a) show the topological edge states located at the interface between (right-handed, 60°, 1.5, 2) and (left-handed, 0°, 2, 1) [see blue arrows]. On the other hand, the eigen displacement fields for the red line in FIG. 5(a) represent the topological edge states located at the interface between (left-handed, 0°, 2, 1) and (right-handed, 60°, 1.5, 2) [see the red arrows in the blue and green cases in FIG. 5(b)].

We note again that the supercell in FIG. 5 is composed of the two types of spiral architectures that went through three times of topological flipping between them (i.e., chirality, rotation angle, and number of turn changes). In APPENDIX C, we also present the projected band structures of the supercells with (i) one-time topological flipping by chirality-only change [i.e., (right-handed, 60°, 1.5, 2) and (left-handed, 60°, 1.5, 2)] and (ii) two-time topological flipping by chirality and rotation angle change [i.e., (right-handed, 60°, 1.5, 2) and (left-handed, 0°, 1.5, 2)]. We confirm that as predicted and guided by our parametric studies Section-II, one(two)-time flipping results in the domain boundary with distinctive (identical) topological nature. We demonstrate that this, in turn, causes the appearance (disappearance) of the topological interface states (see APPENDIX C for details).

Returning to the configuration in FIG. 5, we now explore the topologically protected transport of elastic waves in this elastic metamaterial. We construct the two-part metamaterial consisting of (right-handed, 60°, 1.5, 2) at bottom left and top right, and (left-handed, 0°, 2, 1) at bottom right and top left, as shown in FIG. 5(c). One of the most important features of the edge states in the topological valley metamaterials lies in the valley-polarized nature. In our case, when a vibration source with the frequency of 143 Hz (marked in the green dashed line in FIG. 5(a)) is set on the right side of our elastic metamaterial, the topological valley edge states between (left-handed, 0°, 2, 1) and (right-handed, 60°, 1.5, 2) are excited. As shown in FIG. 5(c), the elastic wave travels along the path at the beginning and when it arrives at the intersection, it propagates both upwards and downwards, but it does not propagate forwards. It indicates that the mode generated at the beginning can couple with the upward mode and the downward mode. We should notice that there is a sharp corner in the upward direction and an obtuse corner in the downward direction, which demonstrates the reflection immunity of topological valley edge states to the path bending. The cross-waveguide splitter also demonstrates the valley-polarized elastic wave propagation [21–25]. The generated forward elastic wave is projected by the K' valley according to the group velocity shown in the projected band structure (FIG. 5(a)). The propagation direction of K(K') valley-projected elastic waves are marked in FIG. 5(c) by yellow(cyan) arrows. Therefore, the elastic wave will travel along the cyan arrows, which forms an elastic wave splitter resulting from its valley-polarization.

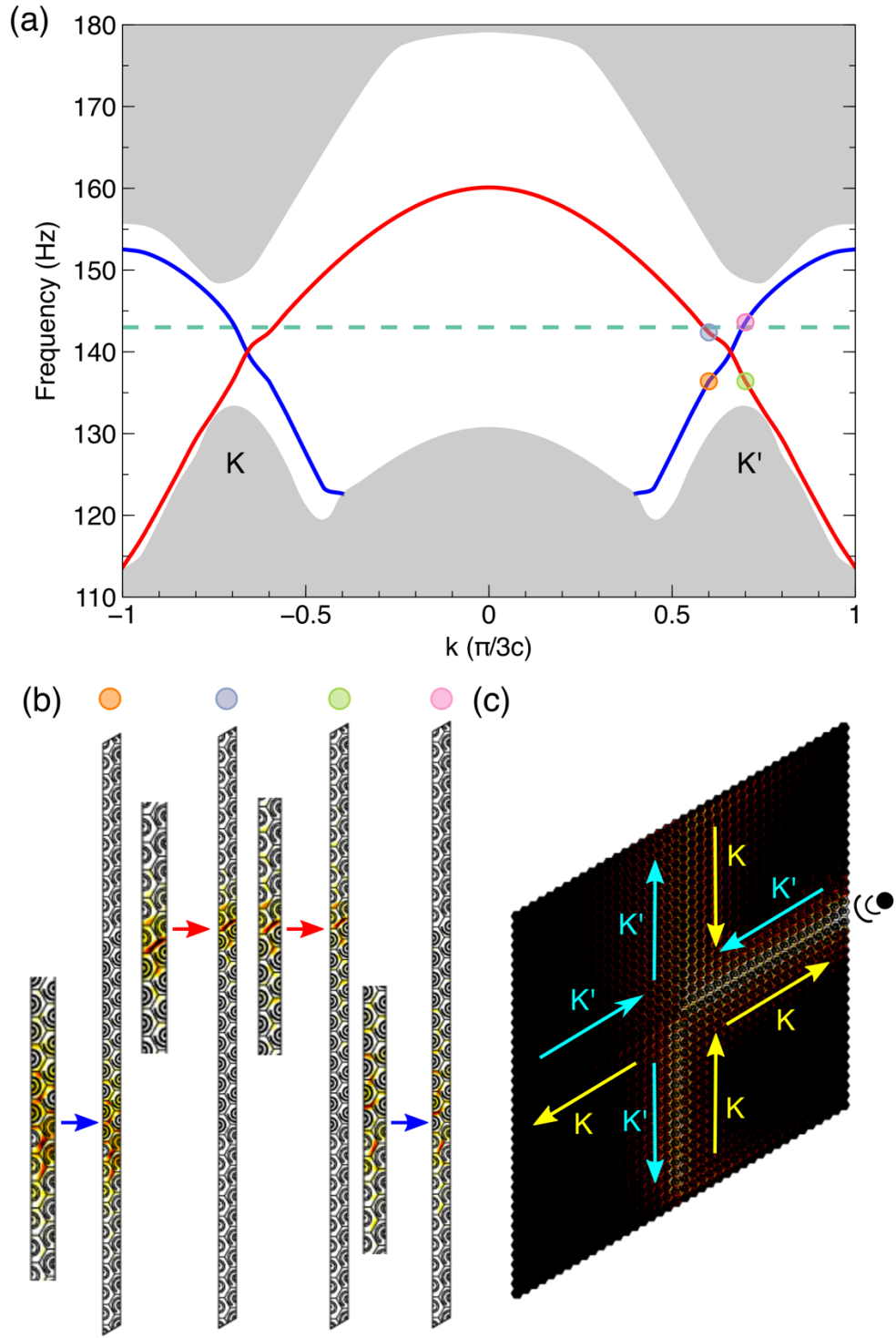

FIG. 5. (a) The projected band structure of the (right-handed, 60, 1.5, 2 | left-handed, 0, 2, 1) metamaterial. Topological interface states are shown in red and blue lines. Bulk modes are shown in gray. Four markers are placed at **k** = 0.6 and **k** = 0.7, respectively. (b) The displacement fields

corresponding to markers in (a). The arrows indicate the position of the interface. The zoom-in displacement fields near interfaces are shown beside the arrows. (c) The transport of elastic waves along the interface. At the frequency of 143 Hz, which is shown in green dashed line in (a), elastic waves propagate along the interface and through the bend.

## IV. CONCLUSIONS

In this study, we investigated topological properties of a spiral structure with multiple degrees of freedom in detail. We demonstrated computationally that the inherent geometrical parameters of the spiral architecture, such as chirality, rotation angle, and the number of turns, can induce the evolution of Berry curvature and valley Chern number of the spiral elastic valley metamaterial, thereby resulting in the landscape change of the topological characteristics. Knowledge of different roles played by spiral parameters enhances the understanding to the elastic topological phase transition using parameters modulation in our system and opens avenues for topological state manipulation. This strategy based on asymmetric spiral architecture gives us the possibility to realize and control topological interface state in a more controllable and efficient manner compared to symmetrical valley architectures. Our research may not be limited only to the monofilar spiral, but can be extended and generalized to the bifilar spiral, trifilar spiral and so on [26], which may have more degrees of freedom to tune the topological properties of elastic metamaterials.

## APPENDIX A: BAND EVOLUTION AT K POINT AND EIGEN MODES

Because of the asymmetric spiral in the metamaterials, when the spiral rotates around the center, there is no degenerate point appearing at K point, though topological phase indeed inverts according to FIG. 2. FIG. 6(a) presents the evolution of two bands at K point as the spiral rotates around the center. It appears the periodic change of frequency does not have degenerate point at K point in this process as we predict. As mentioned in the main text, the eigen modes at K point do not show the inversion of band. The eigen modes of the third and the second bands of (right-handed, 22°, 2, 2) are shown in FIG. 6(b) and FIG. 6(d). For comparison, the eigen modes of the third and the second bands of (right-handed, 38°, 2, 2) are shown in FIG. 6(c) and FIG. 6(e). From the eigen mode, we cannot see any indication of band inversion or topological transition. However, from the Berry curvature in FIG. 2(c) and FIG. 2(e) these two configurations are topologically different. Therefore, exploring the topological property by the direct calculation of Berry curvature is necessary for our study.

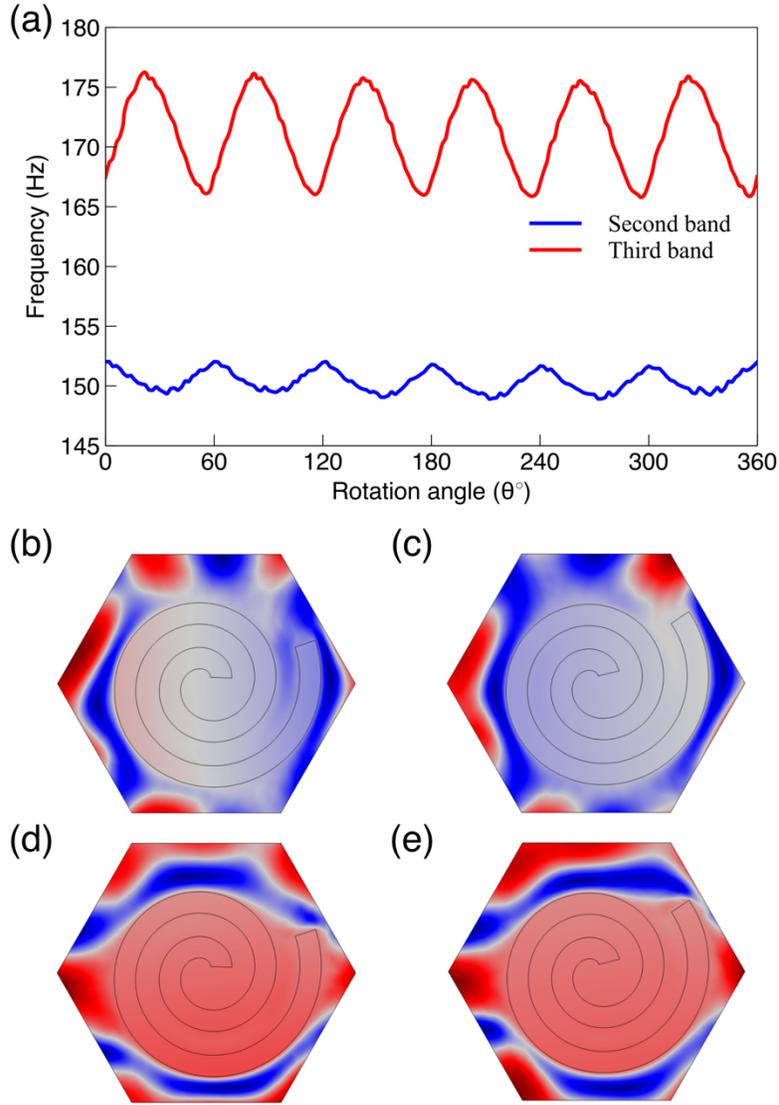

FIG. 6. (a) The blue line represents the evolution of the second band at K point as a function of rotation angle, while the red line represents the evolution of the third band at K point as a function of rotation angle. (b) – (c) The eigen displacement fields of the third bands of (right-handed, 22°, 2, 2) and (right-handed, 38°, 2, 2), respectively. (d) – (e) The eigen displacement fields of the second bands of (right-handed, 22°, 2, 2) and (right-handed, 38°, 2, 2).

**APPENDIX B: BAND GAP EVOLUTION AS A FUNCTION OF THICKNESS**

After we find our desired topological phase of the spiral elastic metamaterials via parameters modulation of chirality, rotation angle and number of turns, we need to find the overlapping band gap between two metamaterials. Tuning thickness $d$ is able to adjust the band gap range. As shown in FIG. 7, red and purple areas are the variations of band gap ranges of (left-handed, 0°, 1.5, $d$) and (left-handed, 0°, 2, $d$), respectively. Both show a clear trend that with the increase of thickness of spiral, the frequency increases and the band gap slightly increases. In this way, we can ensure that two metamaterials we choose have an overlapping band gap.

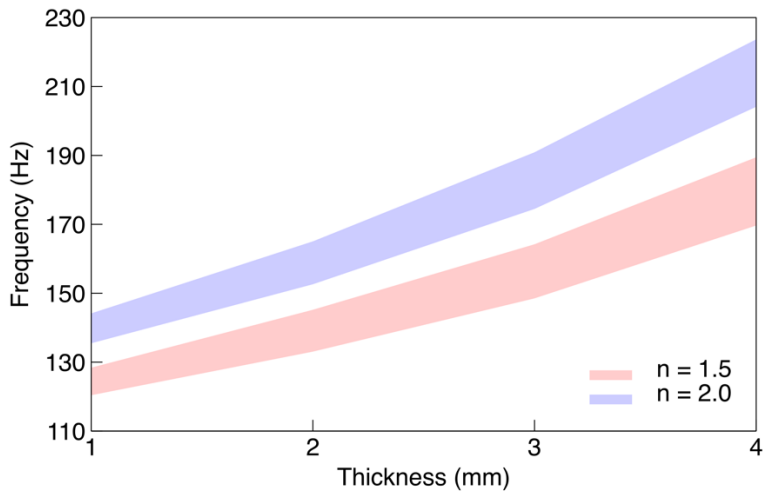

FIG. 7. The variation of band gap range as a function for (left-handed, 0°, 1.5, $d$) and (left-handed, 0°, 2, $d$), when $d$ varies from 1 mm to 4 mm.

## APPENDIX C: PROJECTED BAND STRUCTURE EVOLUTION VIA PARAMETER MODULATION

To demonstrate the topological transition via parameter modulations, we change one parameter at a time. As mentioned in the main text, we choose (right-handed, 60°, 1.5, 2) as one part of the topological insulator. As for the other part, we first change the chirality from right-handed to left-handed. The left panel of FIG. 8 shows the projected band calculated by a supercell (right-handed,

60°, 1.5, 2 | left-handed, 60°, 1.5, 2 | right-handed, 60°, 1.5, 2). There are two topological interface states within the bandgap shown in red and blue, which are located at two interfaces respectively. Next we continue to change the rotation angle to 0°. This operation, along with the previous chirality change, would cause two-time flipping of the topological nature, thereby returning the topological phase to the original one. The corresponding projected band structure is shown in the right panel of FIG. 8. As predicted by our general scheme, (right-handed, 60°, 1.5, 2) and (left-handed, 0°, 1.5, 2) share the same topological phase, resulting in the complete stop band within the bulk band.

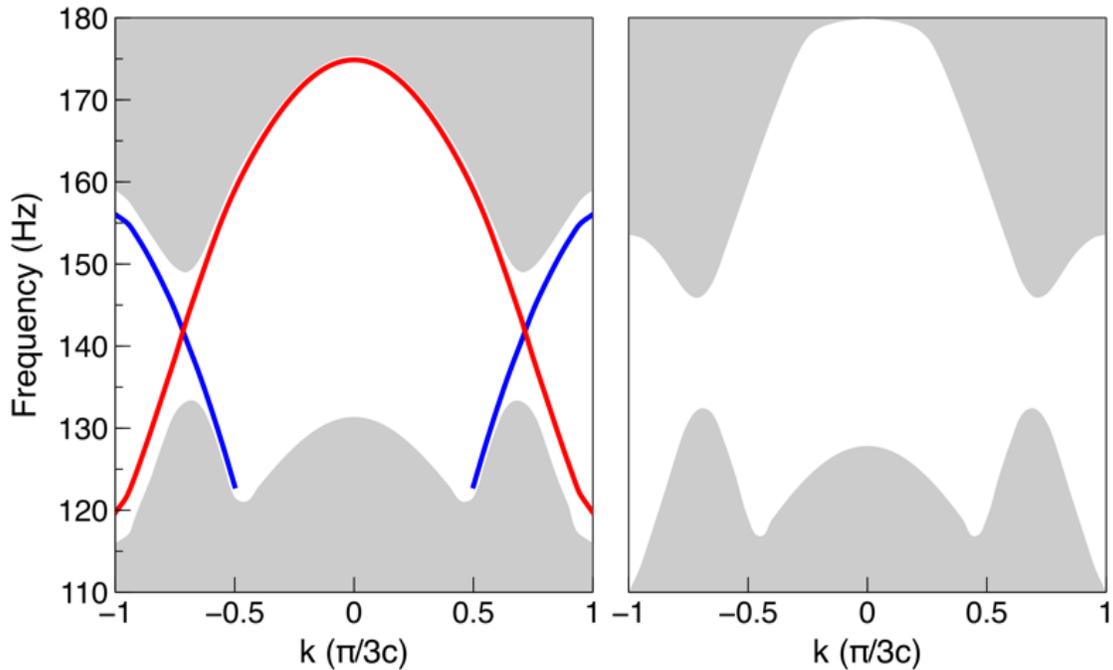

FIG. 8. The left panel shows the projected band structure calculated by a supercell (right-handed, 0°, 2, 1 | left-handed, 0°, 2, 1 | right-handed, 0°, 2, 1). The right panel shows the projected band structure calculated by a supercell (right-handed, 60°, 2, 1 | left-handed, 0°, 2, 1 | right-handed, 60°, 2, 1).

# REFERENCES

[1] A. Rycerz, J. Tworzydło, and C. W. J. Beenakker, *Valley Filter and Valley Valve in Graphene*, Nat. Phys. **3**, 172 (2007).

[2] D. Xiao, W. Yao, and Q. Niu, *Valley-Contrasting Physics in Graphene: Magnetic Moment and Topological Transport*, Phys. Rev. Lett. **99**, 236809 (2007).

[3] K. F. Mak, K. L. McGill, J. Park, and P. L. McEuen, *The Valley Hall Effect in MoS2 Transistors*, Science. **344**, 1489 (2014).

[4] J. W. Dong, X. D. Chen, H. Zhu, Y. Wang, and X. Zhang, *Valley Photonic Crystals for Control of Spin and Topology*, Nat. Mater. **16**, 298 (2017).

[5] X. D. Chen, W. M. Deng, J. C. Lu, and J. W. Dong, *Valley-Controlled Propagation of Pseudospin States in Bulk Metacrystal Waveguides*, Phys. Rev. B **97**, 184201 (2018).

[6] J. Lu, C. Qiu, M. Ke, and Z. Liu, *Valley Vortex States in Sonic Crystals*, Phys. Rev. Lett. **116**, 093901 (2016).

[7] R. K. Pal and M. Ruzzene, *Edge Waves in Plates with Resonators: An Elastic Analogue of the Quantum Valley Hall Effect*, New J. Phys. **19**, 025001 (2017).

[8] J. Vila, R. K. Pal, and M. Ruzzene, *Observation of Topological Valley Modes in an Elastic Hexagonal Lattice*, Phys. Rev. B **96**, 134307 (2017).

[9] S. Li, I. Kim, S. Iwamoto, J. Zang, and J. Yang, *Valley Anisotropy in Elastic Metamaterials*, Phys. Rev. B **100**, 195102 (2019).

[10] X. Xu, W. Yao, D. Xiao, and T. F. Heinz, *Spin and Pseudospins in Layered Transition Metal Dichalcogenides*, Nature Physics.

[11] X. D. Chen, W. M. Deng, J. C. Lu, and J. W. Dong, *Valley-Controlled Propagation of Pseudospin States in Bulk Metacrystal Waveguides*, Phys. Rev. B **97**, 184201 (2018).

[12] L. H. Wu and X. Hu, *Scheme for Achieving a Topological Photonic Crystal by Using Dielectric Material*, Phys. Rev. Lett. **114**, 223901 (2015).

[13] S. Li, D. Zhao, H. Niu, X. Zhu, and J. Zang, *Observation of Elastic Topological States in Soft Materials*, Nat. Commun. **9**, 1 (2018).

[14] T. W. Liu and F. Semperlotti, *Tunable Acoustic Valley-Hall Edge States in Reconfigurable Phononic Elastic Waveguides*, Phys. Rev. Appl. **9**, 014001 (2018).

[15] T. Fukui, Y. Hatsugai, and H. Suzuki, *Chern Numbers in Discretized Brillouin Zone: Efficient Method of Computing (Spin) Hall Conductances*, J. Phys. Soc. Japan **74**, 1674 (2005).

[16] H. Zhu, T. W. Liu, and F. Semperlotti, *Design and Experimental Observation of Valley-Hall Edge States in Diatomic-Graphene-like Elastic Waveguides*, Phys. Rev. B **97**, 174301 (2018).

[17] I. Kim, Y. Arakawa, and S. Iwamoto, *Design of GaAs-Based Valley Phononic Crystals with Multiple Complete Phononic Bandgaps at Ultra-High Frequency*, Appl. Phys. Express **12**, 047001 (2019).

[18] K. Qian, D. J. Apigo, C. Prodan, Y. Barlas, and E. Prodan, *Topology of the Valley-Chern Effect*, Phys. Rev. B **98**, 155138 (2018).

[19] H. Niu, S. Li, and J. Zang, *Reliable and Tunable Elastic Interface States in Soft Metamaterials*, Phys. Status Solidi **14**, 2000338 (2020).

[20] M. Petitjean, *Chirality in Metric Spaces*, Optim. Lett. **14**, 329 (2020).

[21] X. Han, Y. G. Peng, L. Li, Y. Hu, C. Mei, D. G. Zhao, X. F. Zhu, and X. Wang, *Experimental Demonstration of Acoustic Valley Hall Topological Insulators with the Robust Selection of C 3 V-Symmetric Scatterers*, Phys. Rev. Appl. **12**, 14046 (2019).

[22] X. Cheng, C. Jouvaud, X. Ni, S. H. Mousavi, A. Z. Genack, and A. B. Khanikaev, *Robust Reconfigurable Electromagnetic Pathways within a Photonic Topological Insulator*, Nat. Mater. **15**, 542 (2016).

[23] L. Zhang, Y. Yang, M. He, H. Wang, Z. Yang, E. Li, F. Gao, B. Zhang, R. Singh, J. Jiang, and H. Chen, *Valley Kink States and Topological Channel Intersections in Substrate-Integrated Photonic Circuitry*, Laser Photon. Rev. **13**, 1900159 (2019).

[24] B. Z. Xia, T. T. Liu, G. L. Huang, H. Q. Dai, J. R. Jiao, X. G. Zang, D. J. Yu, S. J. Zheng, and J. Liu, *Topological Phononic Insulator with Robust Pseudospin-Dependent Transport*, Phys. Rev. B **96**, 94106 (2017).

[25] X. Wu, Y. Meng, J. Tian, Y. Huang, H. Xiang, D. Han, and W. Wen, *Direct Observation of Valley-Polarized Topological Edge States in Designer Surface Plasmon Crystals*, Nat. Commun. **8**, 1304 (2017).

[26] O. Isik and K. P. Esselle, *Analysis of Spiral Metamaterials by Use of Group Theory*, Metamaterials **3**, 33 (2009).